\documentclass[twocolumn,pre,floatfix]{revtex4}
\usepackage{psfrag,epsfig,amsfonts,amssymb,amsmath,wasysym,bm}
\usepackage{dcolumn}

\usepackage[normalem]{ulem}
\usepackage{color}

\newcommand{\id}{{\rm{I}}_\hr}
\newcommand{\pu}{\tr\{\rho^2\}}
\newcommand{\deff}{d_{\mathrm{eff}}}

\newcommand{\mini}{\mathrm{min}}
\newcommand{\maxi}{\mathrm{max}}
\newcommand{\rhomic}{\rho_{\mathrm{mc}}}
\newcommand{\Amic}{A_{\mathrm{mc}}}
\newcommand{\Imic}{[E-\epsilon,E]}

\newcommand{\hr}{{\cal H}}
\newcommand{\ord}{{\cal O}}

\newcommand{\tr}{\mbox{Tr}}

\providecommand{\norm}[1]{\|#1\|}
\newcommand{\RR}{{\mathbb R}}

\newcommand{\pr}{{\rm{Prob}}}

\begin{document}

\title{Dynamical typicality approach 
to eigenstate thermalization}

\author{Peter Reimann}
\affiliation{Fakult\"at f\"ur Physik, 
Universit\"at Bielefeld, 
33615 Bielefeld, Germany}

\begin{abstract}
We consider the set of all initial states 
within a microcanonical energy shell of
an isolated many-body quantum system,
which exhibit the same, arbitrary 
but fixed non-equilibrium expectation value 
for some given observable $A$.
On condition that this set is not too small,
it is shown by means of a dynamical typicality approach
that most such initial states exhibit thermalization
if and only if $A$ satisfies the so-called weak eigenstate
thermalization hypothesis (wETH).
Here, 
thermalization means that the expectation value
of $A$ spends most of its time close to
the microcanonical value 
after initial transients have died out.
The wETH means that, within the energy shell,
most eigenstates of the pertinent system 
Hamiltonian exhibit very similar expectation values 
of $A$.
\end{abstract}

\maketitle

The eigenstate thermalization hypothesis 
(ETH) plays a pivotal role in numerous recent
investigations of thermalization 
in isolated many-body quantum systems \cite{ale16,gog16},
comparable to the role of the ergodic hypothesis
in the classical realm.
In essence, the ETH postulates that energy eigenstates 
with sufficiently close energy eigenvalues
exhibit very similar expectation values
\cite{deu91,sre94,sre96,rig08}.
It is generally taken for granted that
the ETH guarantees thermalization 
for any initial state with a 
macroscopically well defined system energy.
Whether the ETH is also necessary 
for thermalization is a question of
considerable current interest 
\cite{gog16,tas96,rig12,pal15,tas16,bar17,shi17,mon17}.
Here, we will provide examples implying that ETH 
(in its most common version) should
be considered neither as sufficient nor as necessary 
for thermalization without any further specification 
of the admitted initial states.

Accordingly, we will focus on a suitable 
subset of initial states, namely all pure 
states which exhibit the same, arbitrary 
but fixed initial 
expectation value for some  given
observable $A$.
In the most common case, this
subset is still ``reasonably large''
(in a mathematically precisely defined sense)
and entails quite remarkable dynamical 
typicality and concentration of
measure properties, as detailed in
Refs. \cite{bar09,mul11}.
Here, we further develop 
these concepts and 
show that  a ``weak'' version of the ETH
\cite{bri10,ike13,beu14,iko17,yos17}
is both necessary and sufficient in order that 
the vast majority of those initial states
exhibit thermalization with respect to 
the observable $A$ at hand.
Whether or not a given system thermalizes 
and whether or not it 
satisfies the ETH are very important
issues in themselves, but they are not at the 
focus of our present work. Rather, our main 
focus is on 
how the two issues are connected.

Based on related preliminary conjectures \cite{ber77,vor77},
the ETH was originally proposed in the context of
chaotic systems in the semiclassical limit \cite{sre94,sre96},
see also \cite{shn74,col85,fei85,hel87}.
In fact, for so-called macroscopic observables,
the ETH is already buried in von Neumann's
work \cite{neu29,gol10}, 
as pointed out in Refs.
\cite{gol11,rig12,rei15,ham18}.
More recent analytical investigations of the ETH
often focus on (sums of) local observables,
subsystems in contact with a heat bath,
spatially discrete lattice models, or
Hamiltonians with bound spectra
\cite{tas96,bri10,pal15,tas16,shi17}.
In view of the quite extensive numerical
explorations \cite{ale16,gog16}
and of Deutsch's results 
based on random matrix theory
\cite{deu91,rei15b}, 
this Letter pursues the standpoint
that the ETH is an interesting and relevant
concept beyond any such particular class
of systems and observables.

As usual \cite{ale16,gog16}, 
the isolated many-body system is
described by a Hamiltonian $H$ with
discrete eigenvalues $E_n$ and 
eigenvectors $|n\rangle$.
Focusing on an arbitrary but fixed 
microcanonical energy interval
$\Imic$,
the number of energies $E_n$ 
in this interval
is denoted by $N$ and
we choose the indices so
that $n\in\{1,...,N\}$ for all those $E_n$'s. 
The width $\epsilon$ is assumed to be
small on the macroscopic scale
(well defined system energy) 
but large on the microscopic scale.
For many-body systems with $f\gg 1$ degrees of freedom, 
$N$ is then exponentially large in $f$ \cite{gol10}.
The energy eigenstates $\{|n\rangle\}_{n=1}^N$
span a Hilbert space $\hr$, called the 
microcanonical energy shell.

Considering any given $|\psi\rangle\in\hr$ 
as an initial state $|\psi(0)\rangle$,
it evolves in time according to
$|\psi(t)\rangle=U_t|\psi\rangle$
with $U_t:=e^{-iHt/\hbar}$,
yielding for an arbitrary observable $A$ the 
expectation value
\begin{eqnarray}
& & 
\langle \psi(t)|A|\psi(t)\rangle
=\langle\psi|A_t|\psi\rangle
\ ,
\label{1}
\\
& & A_t := U_t^\dagger A\, U_t 
=
\!\! \sum_{m,n=1}^N\!\! 
A_{mn}\,e^{i(E_m-E_n)t/\hbar}
|m\rangle\langle n|
\ ,
\label{2}
\end{eqnarray}
where $A_{mn}:=\langle m|A|n\rangle$.
In cases where the Hamiltonian $H$ exhibits degeneracies,
its eigenvectors $|n\rangle$ 
are chosen so that 
the matrix $A_{mn}$ is diagonal 
within every eigenspace.
Denoting averages over all times $t\geq 0$ 
by an overbar, it follows that
\begin{eqnarray}
\overline{A_t} =\sum_{n=1}^N
A_{nn}\, |n\rangle\langle n|
\ ,
\label{3}
\end{eqnarray}
and for the time averaged expectation value in (\ref{1}) that
\begin{eqnarray}
A_\psi
:=\overline{\langle \psi(t)|A|\psi(t)\rangle}
=\langle\psi|\overline{A_t}|\psi\rangle
=
\sum_{n=1}^N
|\langle n|\psi\rangle|^2\, A_{nn}
\ .
\label{4}
\end{eqnarray}

The most common or ``strong'' version of ETH 
(sETH) states \cite{ale16,gog16}
that the diagonal matrix elements $A_{nn}$ assume 
very similar values for all $n\in\{1,...,N\}$.
Consequently, the long time average in (\ref{4})
is very well approximated by the microcanonical 
expectation value $\Amic:=\tr\{\rhomic A\}$, where
$\rhomic := \id/N$
and $\id:=\sum_{n=1}^N|n\rangle\langle n|$
(identity on $\hr$).
Since this is precisely the prediction
of textbook statistical mechanics for 
our system at thermal equilibrium,
and since this property applies to
any initial condition $|\psi\rangle\in\hr$,
it is tempting to conclude that 
the sETH implies thermalization.
However, 
one can readily
tailor initial conditions and observables, 
which fulfill the sETH and 
$A_\psi \simeq \Amic$,
while the expectation values in (\ref{1})
maintain non-negligible oscillations
{\em ad infinitum}, 
i.e., they do not exhibit thermalization 
in any meaningful sense.
For example, $|\psi\rangle=(|1\rangle+|2\rangle)/\sqrt{2}$,
$A_{12}=A_{21}=1$, and $A_{mn}=0$ for all other $m,n$
yields $\langle \psi(t)|A|\psi(t)\rangle=\cos(\omega t)$
with $\omega:=(E_2-E_1)/\hbar$.
One may object that this example is
experimentally unrealistic 
\cite{rei08} 
and incompatible
with the generalized ETH postulated in 
Ref. \cite{sre96}, yet there seems to be no argument
which {\em rigorously} disqualifies {\em all} 
counter-examples of this kind.
Accordingly, the sETH should not be 
considered as sufficient for thermalization 
without any further conditions regarding the 
observables or the initial conditions.

Henceforth, we adopt the standard notion of thermalization 
from Refs. \cite{sre96,neu29,gol10,tas96,rei08,lin09}, 
requiring that not only the time averaged, 
but also the instantaneous expectation 
values in (\ref{1}) must be close to
$\Amic$ 
for the vast majority of all sufficiently 
large times $t$, i.e.,
after initial transients have died out.
Note that a small fraction of 
exceptional times $t$ is unavoidable, e.g.,
due to quantum revivals,
caused by the 
quasi-periodicity of $A_t$ in (\ref{2}).
In addition to 
$A_\psi\simeq\Amic$,
we thus require that
\begin{eqnarray}
\overline{\left(\langle\psi|A_t|\psi\rangle-A_\psi\right)^2}
\ll 1
\ .
\label{5}
\end{eqnarray}
As demonstrated, e.g., in Refs.
\cite{rei08,lin09,sho11,rei12,sho12,bal16},
an arbitrary $|\psi\rangle\in\hr$ satisfies
(\ref{5}) under the sufficient condition 
\begin{eqnarray}
S_\psi:= 
\sum_{n=1}^N |\langle n|\psi\rangle|^4\ll 1
\ ,
\label{6}
\end{eqnarray}
where we tacitly
restricted ourselves to the generic case 
\cite{neu29,tas96,rei08,lin09}
that the energy differences
$E_m-E_n$ are finite and mutually 
different for all pairs $m\not=n$
(generalizations are possible
\cite{sho11,rei12,sho12,bal16}
but omitted here for the sake of simplicity).
We thus can conclude that the sETH 
together with (\ref{6}) are sufficient conditions 
for thermalization.

On the other hand, we will later provide examples 
which exhibit thermalization but violate the sETH.
Altogether, the sETH alone is thus neither 
sufficient nor necessary for thermalization:
We have to modify or supplement the 
sETH criterion, or we have to admit exceptions 
and show that they are ``rare'' in some 
suitable sense.
In the following, we work out an approach 
along these lines.

To begin with, we note that the original Hilbert space
of the system is usually much larger
than the energy shell $\hr$, and that $A$ and $H$ are
{\em a priori} operators on that larger space.
Accordingly, $\id:=\sum_{n=1}^N |n\rangle\langle n|$ 
may also be considered as a projector onto $\hr$ 
and $A':=\id A \id$ as the restriction or projection
of $A$ onto $\hr$ (and likewise for $H$).
But since only vectors $|\psi\rangle$ with support 
in $\hr$ are considered in (\ref{1}), one readily sees 
that every single term in (\ref{1})-(\ref{6}) remains
exactly the same if we replace $A$ by $A'$.
In particular, $A_{nn}=A'_{nn}$ for all 
$n\in\{1,...,N\}$.
On the other hand, the eigenvalues and
eigenvectors of $A'$, henceforth denoted as
$a_n$ and $|\varphi_n\rangle$, respectively,
are in general different from those of $A$.
From now on, we always work with $A'$
but -- for the sake of convenience and since 
it actually does not matter in most formulas --
we again omit the prime symbol.

Possibly after adding a trivial constant 
to the observable and multiplying it by a 
constant factor, we can and will assume that
\begin{eqnarray}
\tr\{A\} & = & 0
\ ,
\label{7}
\\
\norm{A} & = & 1
\ ,
\label{8}
\end{eqnarray}
where $\tr\{\cdot\}$ is the trace in $\hr$ and 
$\norm{\cdot}$ the operator norm.
It follows that $a_{\maxi}:=\max_na_n>0$
and $a_{\mini}:=\min_n a_n<0$.
For an arbitrary but fixed
$a\in(0,a_{\maxi})$, we define
\begin{eqnarray}
g(x) & := & \frac{1}{N}\sum_{n=1}^N \frac{1}{1+x(a-a_n)}
\ .
\label{9}
\end{eqnarray}
One readily verifies that
$g(0)=1$, $g'(0)=-a<0$, $g(x)\to\infty$ as $x$ approaches
$x_{\maxi}:=1/(a_{\maxi}-a)$ from below, 
and $g''(x)>0$ for all $x\in[0,x_{\maxi})$.
These properties imply
that there must be exactly one 
$x\in(0,x_{\maxi})$ with $g(x)=1$.
This $x$ value is henceforth 
denoted as $y(a)$.
One thus can conclude that
$y(a)>0$, that
\begin{eqnarray}
p_n & := & \frac{1}{N}\frac{1}{1+y(a)\,(a-a_n)}>0
\label{10}
\end{eqnarray}
for all $n=1,...,N$, and that
\begin{eqnarray}
\sum_{n=1}^N p_n & = & 1
\ .
\label{11}
\end{eqnarray}
Similarly, for $a\in(a_{\mini},0)$ there is 
a unique $y(a)<0$ which satisfies 
(\ref{10}) and (\ref{11}),
while $y(a)$ must be zero for $a=0$.
Finally, one can deduce from (\ref{10}) and (\ref{11})
by means of a straightforward calculation
\cite{f1} that
\begin{eqnarray}
\sum_{n=1}^N a_n\, p_n & = & a
\label{12}
\end{eqnarray}
for any given $a\in (a_{\mini},a_{\maxi})$.

Next, we introduce an ensemble of 
random vectors $|\chi\rangle\in\hr$ 
via
\begin{eqnarray}
|\chi\rangle=
\sum_{n=1}^N c_n\, |\chi_n\rangle
\ ,
\label{13}
\end{eqnarray}
where $\{|\chi_n\rangle\}_{n=1}^N$ is 
any orthonormal basis of $\hr$,
and where the real and imaginary parts 
of the $c_n$'s are independent, 
Gaussian distributed random variables 
of mean zero and variance $1/2N$.
Denoting averages over the $c_n$'s
by $[...]_c$, it follows that 
$[c^\ast_m c_n]_c=\delta_{mn}/N$
for all $m,n\in\{1,...,N\}$.
The random vectors 
(\ref{13}) are thus normalized on the average,
$[\langle\chi |\chi\rangle]_c=1$,
but not individually.
Moreover,
the random vector 
ensemble is invariant under arbitrary 
unitary transformations of the 
basis $\{|\chi_n\rangle\}_{n=1}^N$
(all its statistical properties 
remain unchanged).
All bases are thus equivalent
and the ensemble is unbiased.
In terms of this ensemble, yet another 
ensemble of random 
vectors $|\phi\rangle$ is defined via
\begin{eqnarray}
|\phi\rangle & := & \sqrt{N}\,\rho^{1/2}\,|\chi\rangle
\label{14}
\ ,
\\
\rho & := & \sum_{n=1}^N p_n\, |\varphi_n\rangle\langle \varphi_n|
\ ,
\label{15}
\end{eqnarray}
where the $|\varphi_n \rangle$ 
have been introduced above (\ref{7})
and where $\rho^{1/2} := \sum_{n=1}^N \sqrt{p_n}\, |\varphi_n\rangle\langle \varphi_n|$, 
implying $(\rho^{1/2})^2=\rho$.
Note that $\rho$ is Hermitian, positive
(see (\ref{10})), and of unit trace (see (\ref{11})),
i.e., a well defined density operator.

Given any Hermitian operator $B:\hr\to\hr$, 
one readily can infer from (\ref{13})-(\ref{15}) 
that \cite{f2}
\begin{eqnarray}
\mu_B & := & [\langle\phi|B|\phi\rangle]_c
= \tr\{\rho  B\}
\ ,
\label{16}
\\
\sigma_B^2 & := &
\left[
\left(\langle\phi|B|\phi\rangle
- \mu_B
\right)^2
\right]_c
=
 \tr\{(\rho  B)^2\} 
 \ .
\label{17}
\end{eqnarray}
Taking advantage of the Cauchy-Schwarz inequality \cite{f3},
$ \tr\{(\rho  B)^2\}$ can be upper bounded by
 $\tr\{\rho^2  B^2\}$.
Evaluating the trace by means of the 
eigenbasis of $B$, one thus obtains
\begin{eqnarray}
\sigma_B^2 \leq \norm{B}^2\,  \pu 
\ .
\label{18}
\end{eqnarray}

In the following, we restrict 
ourselves to the case
\begin{eqnarray}
\deff:=1/\pu  \gg 1 \ .
\label{19}
\end{eqnarray}
As observed in \cite{lin09},
the effective dimension $\deff$
tells us, how many pure 
states contribute appreciably to the 
mixture $\rho$.
Indeed, 
one readily finds -- similarly 
as in footnote \cite{f2} -- that
$[\,|\phi\rangle\!\langle\phi |\,]_c=\rho$.
Moreover, if $p_n=1/M$ for $M$ of the
weights $p_n$ in (\ref{15}),
then $\deff=M$, and the $|\phi\rangle$ in (\ref{14}) 
arise by unbiased sampling of vectors within an 
$M$ dimensional subspace of $\hr$.
In other words,
$\deff$ quantifies the
``diversity'' of random vectors 
$|\phi\rangle$ contributing to $\rho$,
and (\ref{19}) ensures that
the ensemble of random vectors 
in (\ref{14}) is 
not ``too small''.
Moreover, it is reasonable to expect 
that, unless $a$ is very close to $a_{\maxi}$
or $a_{\mini}$, many $p_n$'s will
notably contribute in (\ref{12}), and 
hence the effective dimension of 
$\rho$ will be large.
This expectation is quantitatively
confirmed in the 
Supplemental Material \cite{sup}, 
showing 
that $\deff$ is in fact exponentially 
large in the system's degrees of 
freedom under quite general conditions.

For $B=\id$,
it follows from (\ref{16})-(\ref{19}) that
$[\langle\phi|\phi\rangle]_c = 1$
and 
$[\left(\langle\phi|\phi\rangle- 1 \right)^2]_c \ll 1$.
The vast majority of all  $|\phi\rangle$ in (\ref{14}) 
thus exhibit norms very close 
to unity.
Next, by choosing $B=A$
it follows with (\ref{12}),
(\ref{15}), and (\ref{16}) that
$\mu_{\! A}= a$
and with (\ref{8}), (\ref{17})-(\ref{19}) that
$\sigma_A^2\ll 1$.
The vast majority of all  
$|\phi\rangle$ in (\ref{14}) 
thus exhibit expectation 
values $\langle\phi|A|\phi\rangle$ 
very close to the preset value $a$.
Likewise, by choosing $B=\overline{A_t}$ 
and observing that 
$\norm{\overline{A_t}}\leq\norm{A}$, one can infer
from (\ref{8}) and (\ref{17})-(\ref{19}) that the vast 
majority of all $|\phi\rangle$ in (\ref{14}) 
yield time averaged expectation values 
$A_\phi$ in (\ref{4})
very close to 
$\tr\{\rho \overline{A_t}\}$. 
Finally, one can show by similar 
calculations as in footnote 
\cite{f2} that
$S_\phi$ from (\ref{6})
satisfies $[S_\phi]_c\leq 2/\deff$.
Observing (\ref{19}) and $S_\phi\geq 0$
it follows that $S_\phi$ 
must be very small for most 
$|\phi\rangle$'s from (\ref{14}).

So far, the initial states $|\phi\rangle$ 
in (\ref{14}) are in general not 
normalized. But, as seen above, 
the vast majority among them are
almost of unit length.
Hence, if we replace for every given
$|\chi\rangle$ the concomitant
$|\phi\rangle$ in (\ref{14}) by 
its strictly normalized counterpart
\begin{eqnarray}
|\psi\rangle := 
\langle\chi|\rho |\chi\rangle^{-1/2}\,
\rho^{1/2}|\chi\rangle
\label{20}
\end{eqnarray}
then the ``new'' expectation values
$\langle\psi|A|\psi\rangle$
and
$\langle\psi|\overline{A_t}|\psi\rangle$ 
will mostly remain very close to
the ``old'' ones, i.e., to
$\langle\phi|A|\phi\rangle$
and 
$\langle\phi|\overline{A_t}|\phi\rangle$,
respectively.
Likewise, $S_\psi$ must remain very 
small for most $|\psi\rangle$'s.
More precisely, one can show \cite{sup} that a
vector $|\psi\rangle$, 
randomly sampled according 
to (\ref{13}) and (\ref{20}),
satisfies simultaneously the 
three conditions
$| \langle\psi|A|\psi\rangle - a |\leq 2 \delta$,
$| \langle\psi| \overline{A_t} |\psi\rangle - \tr\{\rho \overline{A_t}\} | \leq 2 \delta$,
and
$S_\psi \leq 4 \delta$
with probability $P\geq1-6\delta$,
where $\delta:=\deff^{-1/3}$
is exponentially 
small in the system's degrees of 
freedom.

In conclusion, 
the vast majority of all
initial states $|\psi(0)\rangle:=|\psi\rangle$
from (\ref{20})
exhibit initial expectation values 
$\langle \psi (0)|A| \psi (0)\rangle$
very close the preset value $a$
in (\ref{9}), (\ref{10}), and
the time average in (\ref{4})
satisfies very well the approximation
\begin{eqnarray}
A_\psi
=\tr\{\rho \overline{A_t}\} 
\ .
\label{21}
\end{eqnarray}
In other words, the long time limit (\ref{4})
is for most $|\psi\rangle$ 
very close to one and the same value, 
given by the right hand side 
of (\ref{21}).
As discussed below (\ref{4}), we furthermore
require as a necessary condition for 
thermalization that those very 
similar long time averages of most 
$|\psi\rangle$'s must be close to the 
microcanonical expectation value $\Amic$.
Exploiting (\ref{3}) to infer
$\Amic=\tr\{\rhomic \overline{A_t}\}$
\cite{f4},
it follows that
the right hand side of (\ref{21}) must satisfy
\begin{eqnarray}
\tr\{\rho \overline{A_t}\} =
\tr\{\rhomic \overline{A_t}\}
\label{22}
\end{eqnarray}
in very good approximation.
Recalling that under the same premise (\ref{19})
most $|\psi\rangle$'s also satisfy (\ref{6}),
we can conclude that
(\ref{19}) and (\ref{22}) are sufficient
to guarantee that most $|\psi\rangle$'s
from (\ref{20}) exhibit thermalization.

The main feature of the random vector
ensemble (\ref{20}) is that the expectation 
value $\langle\psi|A|\psi\rangle$ 
is {\em almost} equal to 
$a$ for {\em most} 
$|\psi\rangle$'s.
As can be inferred from Ref. \cite{mul11},
this ensemble yields  results for the 
statistics (mean and variance) of 
$A_\psi$ and $S_\psi$
which are  very similar to those for an ensemble,
where {\em all} normalized 
vectors, whose expectation value is 
{\em strictly} equal to $a$, are 
realized with equal probability 
(and all other vectors are excluded).
We thus can conclude that
most initial states 
$|\psi\rangle\in\hr$ with 
$\langle\psi|A|\psi\rangle=a$
exhibit thermalization, provided
(\ref{19}) and (\ref{22}) 
are fulfilled.

In principle, the observable 
$A$ and the value of  $a$
uniquely determine $y(a)$ in
(\ref{10}) and (\ref{11}). 
Hence, $\rho$ in (\ref{15}) follows 
and condition
(\ref{22}) can be checked.
In practice, a general, explicit 
solution of all the necessary 
equations seems not possible.
We thus content ourselves with
a series expansion in powers of 
$a$.
Since $y(0)=0$ (see below (\ref{11})),
we can expand $y(a)$ as $y'(0)a+y''(0)a^2/2+...$
and the denominator in (\ref{10})
as a geometric series.
Substituting all this into (\ref{11})
and comparing terms with equal 
powers of $a$ yields equations
for 
$y'(0)$, $y''(0)$,...
which can be iteratively solved.
As a result, Eq. (\ref{15}) assumes the form
\begin{eqnarray}
\rho & = & \rhomic + 
\frac{1}{N} \sum_{k=1}^\infty [y(a)(A-a)]^k
\ ,
\label{23}
\\
y(a) & = & (1/m_2)\, a  -\ (m_3/m_2^3)\, a^2 + 
\ord(a^3)
\ ,
\label{24}
\\
m_k & := & \frac{1}{N}\sum_{n=1}^N(a_n)^k=\tr\{\rhomic A^k\}
\label{25}
\ .
\end{eqnarray}
Taking into account Eq. (\ref{3}), this finally yields
\begin{eqnarray}
\tr\{\rho \overline{A_t}\} - 
\tr\{\rhomic \overline{A_t}\}
=
a\sum_{n=1}^N \frac{(A_{nn})^2}{m_2N}
+ \ord(a^2)
\ .
\label{26}
\end{eqnarray}
In view of the approximation (\ref{22}), 
the coefficients on the right hand side 
of (\ref{26}) must be zero 
(or very small) separately 
for every power of $a$.
Together with (\ref{25}) we thus
can conclude that
\begin{eqnarray}
\frac{1}{N}\sum_{n=1}^N (A_{nn})^2\ll 
\tr\{\rhomic A^2\}\leq 1
\ ,
\label{27}
\end{eqnarray}
where we utilized (\ref{8}) in the last step.
This is the main result of our paper.
It implies that {\em most} $A_{nn}$'s
must be very small \cite{f4}.
In other words, the values
of $\langle n|A|n\rangle$
must be very similar to each other
for most energy eigenvectors $|n\rangle$
with eigenvalues $E_n$ in the considered 
energy interval $[E-\epsilon, E]$.
Following Refs. \cite{bri10,ike13,beu14,iko17,yos17},
the latter property is denoted 
as the {\em weak ETH} (wETH).
In Ref. \cite{bar17}, somewhat
similar results have been 
obtained for some particular initial (mixed)
states which arise by certain, 
very small perturbation of a canonical 
density operator \cite{barPC}.

In short, we found that typicality of thermalization 
implies the wETH.
In the opposite case, i.e., when most $|\psi\rangle$'s 
do not exhibit thermalization, then most of them still
approach very similar long time averages
according to (\ref{21}).
However, (\ref{22}) is no longer 
fulfilled, hence the right hand side of (\ref{26})
is non-negligible and the wETH is violated.
In other words, wETH implies typicality of 
thermalization.
As announced below (\ref{6}),
a system which violates the sETH thus exhibits
thermalization provided it still satisfies the wETH.
Moreover, it is noteworthy that 
-- at least for not too large $a$ values --
the typical deviation from the thermal 
expectation value 
$\Amic=\tr\{\rhomic \overline{A_t}\}=0$
\cite{f4} in (\ref{26}) exhibits the same 
sign as the initial expectation 
value $\langle\psi |A| \psi \rangle=a$
itself.

Clearly, in all those conclusions, 
Eq. (\ref{26}) plays a pivotal role, 
connecting the decisive quantity for 
thermalization (left hand side) with the 
essential quantifier of wETH (sum on the right hand side).
Our above line of reasoning thus 
has the virtue of being
concise and ``natural''.
Its shortcoming is that the arguments
are not mathematically rigorous.
(In fact, already the convergence of the expansions
in (\ref{23}) and (\ref{24}) may strictly speaking
be questionable.)
A complementary, 
more rigorous but 
less enlightening line of reasoning 
is provided as Supplemental 
Material \cite{sup}.

In conclusion, the weak ETH 
has been established as a 
necessary and sufficient
prerequisite for thermalization in 
isolated many-body quantum systems 
in the following sense:
The vast majority of all pure states, 
which exhibit the same initial expectation 
value for some observable $A$,
closely approach the pertinent
microcanonical expectation value 
of $A$ for practically all 
sufficiently large times.
It is remarkable that also 
in several other related studies
it is the weak rather than the
strong ETH which naturally arises
\cite{col85,hel87,bar17,ste14}.
Note that the necessity of the (weak or strong) ETH
for thermalization is not something
that one might have expected {\em a priori}
due to some intuitively quite obvious 
reasons \cite{gog16,pal15}.
For instance, Peres argues \cite{per84}
that generic (chaotic) systems 
should entail pseudorandom $A_{nn}$'s,
which are statistically independent 
of the $|\langle n|\psi \rangle|^2$
in (\ref{4})
for most $|\psi\rangle$.
If this quite reasonable looking expectation 
was correct, then
the right hand side of (\ref{4})
could be well approximated by 
$\tr\{\rhomic A\}$, implying thermalization
even if the (weak or strong)
ETH were violated.
In contrast, our key relation (\ref{26}) 
shows that the $|\langle n|\psi \rangle|^2$ 
and the $A_{nn}$ in (\ref{4}) must 
be ``correlated'' in a very subtle manner, 
except for the ``trivial case'' that most 
of the $A_{nn}$'s are very similar to each 
other, i.e., unless $A$ satisfies the 
weak ETH in the first place.
Put differently, whenever typical non-equilibrium 
initial states do not exhibit thermalization, 
then such correlations must be a generic 
feature.
Indeed, they can be seen in numerical
examples \cite{rig08}, 
but their intuitive physical origin 
previously appeared to be a mystery 
to the present author.
Our dynamical typicality 
approach provides at least
a first step towards its resolution:
In order to exhibit any non-thermal 
expectation value, most initial 
states $|\psi\rangle$ in (\ref{20})
must necessarily acquire some sort of
``correlation'' with $A$ via (\ref{10}) and
(\ref{15}).

\begin{acknowledgments}
Inspiring discussions with Jochen Gemmer, 
Ben N. Balz, Christian Bartsch,
and Lennart Dabelow 
are gratefully acknowledged.
This work was supported by the 
Deutsche Forschungsgemeinschaft (DFG)
under Grant No. RE 1344/10-1 and
within the Research Unit FOR 2692
under Grant No. RE 1344/12-1.
\end{acknowledgments}

\newpage
{\color{white}{.}}
\\[20cm]
{\color{white}{.}}
\newpage

\begin{center}
{\bf{\large{SUPPLEMENTAL MATERIAL}}}
\end{center}

Throughout this Supplemental Material, equations
from the main paper are indicated by an extra letter 
``$m$''. For example, ``Eq. (m1)'' refers to 
equation (1) in the main paper.

%%%%%%%%%%%%%%%%%%%%%%%%%%%%%%%%%%%%%%%%%
\section{Large effective dimensions}
\label{s1}
In this section we substantiate the assertion
below (m19) 
that $\deff$ is exponentially large in the 
system's degrees of freedom under quite 
weak assumptions about $A$
and with the possible
exception of $a$ values very close 
to $a_{\maxi}$ or $a_{\mini}$.

We recall that the domain of
$a$ values admitted in (m9)
is given by $(a_{\mini},a_{\maxi})$,
hence $a_{\maxi}-a_{\mini}$
may be viewed as the range of $a$.
With (m7) and (m8) it follows 
that $a_{\maxi}$ is positive, $a_{\mini}$
is negative, and at least one of them 
is of unit modulus, implying that
\begin{eqnarray}
1< a_{\maxi}-a_{\mini} \leq 2
\ .
\label{a1}
\end{eqnarray}

We also recall that once the 
observable $A$
and the value of 
$a\in(a_{\mini},a_{\maxi})$ in (m9)
are fixed, there is a unique
$y(a)$ in (m10) which satisfies 
(m11) and (m12).
Furthermore, we can and will 
restrict ourselves to the case
\begin{eqnarray}
0\leq a < a_{\maxi} \ ,
\label{a2}
\end{eqnarray}
since the corresponding results for 
$a_{\mini} < a < 0$ then readily follow
by considering $-A$ instead of $A$.
Finally, we recall that
\begin{eqnarray}
y(a)\geq 0
\label{a3}
\end{eqnarray}
within the domain (\ref{a2}), see
above (m10).

Defining the Heaviside step function as
$\Theta (x):=1$ for $x > 0$ and $\Theta(x)=0$
for $x\leq 0$, the fraction (relative number) of 
eigenvalues $a_n$ greater than $a$ is given by
\begin{eqnarray}
\nu_a :=\frac{1}{N}\sum_{n=1}^N \Theta(a_n-a)
\ .
\label{a4}
\end{eqnarray}
Moreover,  we denote by $\tilde a$ the largest $a$ 
value which satisfies $\nu_a\geq 2/\ln N$.
Observing that $\nu_a$, considered as a function 
of $a$, increases in steps of $1/N$, 
and focusing on large $N$ (see below), we can 
conclude that in very good approximation
\begin{eqnarray}
\nu_{\tilde a} = 2/\ln N
\ .
\label{a5}
\end{eqnarray}

We are now in the position to formulate our 
assumptions regarding the observable $A$:
The spectrum of $A$ 
is supposed to exhibit an approximately 
constant and not too small density of eigenvalues 
within the domain $[\tilde a,a_{\maxi}]$.
More precisely, when rewriting 
(m10), (m11), and (\ref{a4})
for $a=\tilde a$ as
\begin{eqnarray}
& & 
\int^{a_{\maxi}}_{a_{\mini}} 
dx \ w(x)\, 
\frac{1}{1+y(\tilde a)\,(\tilde a-x)}=1
\ ,
\label{a6}
\\
& & 
\int^{a_{\maxi}}_{a_{\mini}}
dx \ w(x)\, \Theta(x-\tilde a) = \nu_{\tilde a}
\ ,
\label{a7}
\\
& & 
w(x) := \frac{1}{N} \sum_{n=1}^N \delta(a_n-x)
\ ,
\label{a8}
\end{eqnarray}
we assume that 
$w(x)$ in (\ref{a6}) and (\ref{a7})
can be approximated reasonably 
well as
\begin{eqnarray}
w(x) = \tilde w  \ \mbox{for all $x\in[\tilde a, a_{\maxi}]\,$,}
\label{a9}
\end{eqnarray}
for some suitably chosen constant $\tilde w$,
which furthermore satisfies the condition
\begin{eqnarray}
\tilde w  \gg 2/\ln N
\ .
\label{a10}
\end{eqnarray}
Note that the function $w(x)$ in (\ref{a8})
is normalized to unity and thus may 
be viewed as an eigenvalue probability 
distribution.

Given the above assumptions (\ref{a9}), (\ref{a10})
are fulfilled,
one can infer from (\ref{a7}), (\ref{a9}) 
the approximation
\begin{eqnarray}
\nu_{\tilde a} = (a_{\maxi}-\tilde a)\, \tilde w   
\,
\label{a11}
\end{eqnarray}
and with (\ref{a5}), (\ref{a10}) it follows that
\begin{eqnarray}
0 < a_{\maxi}-\tilde a\ll 1
\ .
\label{a12}
\end{eqnarray}
Likewise, one can show that 
(\ref{a6}), (\ref{a9}), (\ref{a10}) imply
\begin{eqnarray}
N \geq \deff\geq \sqrt{N}\ \mbox{for all $a\in[0,\tilde a]\,$.}
\label{a13}
\end{eqnarray}
Before providing the detailed derivation of
this result, we first turn to its discussion.

As mentioned above Eq. (m1),
for systems with $f\gg 1$ degrees of 
freedom, $N$ is exponentially large in $f$.
It follows that the effective dimension 
in  (\ref{a13}) is exponentially 
large in $f$ as well.
Moreover, the right hand side of
(\ref{a10}) will be roughly comparable in 
order of magnitude to $1/f$.
For macroscopic systems with, say,
$f=\ord(10^{23})$ degrees of
freedom, this means that 
$\deff$ in (\ref{a13}) must be
unimaginably large, and that 
the right hand side of (\ref{a10})
must be extremely small.

From (\ref{a1}) and (\ref{a12}) 
we can conclude that $\tilde a$
is very close to $a_{\maxi}$
compared to the full range 
$a_{\maxi}-a_{\mini}$ of admitted 
$a$ values.
Hence, the vicinity of $a_{\maxi}$
excluded in (\ref{a13}) is very small.
Likewise, the interval $[\tilde a,a_{\maxi}]$
in (\ref{a9}) is very small.
Nevertheless, the number of eigenvalues
$a_n$ contained in this interval is very
large, namely $2N/\ln N$, as 
can be deduced from (\ref{a4}) 
and (\ref{a5}).
Therefore, approximating the 
eigenvalue probability distribution $w(x)$ 
from (\ref{a8})
within the small interval $[\tilde a,a_{\maxi}]$
by some constant value $\tilde w$
is expected to be possible under 
quite general conditions.
In other words, our assumption (\ref{a9})
will be satisfied by a quite large 
class of observables $A$.

Next we remark that the
global mean value of $w(x)$, i.e. 
the average of (\ref{a8}) over all
$x\in [a_{\mini},a_{\maxi}]$, is given
by $w_{av}=1/(a_{\maxi}-a_{\mini})$.
With (\ref{a1}) it follows that 
$w_{av}>1/2$.
Compared to this global average
value $w_{av}$, the condition (\ref{a10})
on the local average $\tilde w$
from (\ref{a9})  is very weak.
In other words, also our second 
assumption (\ref{a10}) is expected to be 
satisfied by a quite large class of 
observables $A$.

Recalling that analogous conclusions 
apply to $a<0$ (see below (\ref{a2})),
we thus recover the assertion below 
(m19).
Moreover, similar results can
also be derived for observables
$A$ with other types of spectral properties near 
$a_{\maxi}$ and $a_{\mini}$ [1].

Finally, we turn to the derivation of the two 
inequalities in (\ref{a13}).
The first inequality readily follows from the 
well known fact that the purity $\pu$ in (m19) 
is minimized by the microcanonical ensemble 
$\rhomic$. We are thus left with 
the second inequality.

By exploiting (m10), (m11), 
(m15), and (\ref{a3}) we can conclude
\begin{eqnarray}
\pu
& = & 
\sum_{n=1}^N p_n^2\leq \sum_{n=1}^N p_n p_{\maxi}=p_{\maxi}
\ ,
\label{a14}
\\
p_{\maxi} 
& := & 
\max_{n} p_n = \frac{1}{N}\frac{1}{1-y(a)(a_{\maxi}-a)}
\ .
\label{a15}
\end{eqnarray}
Together with the definition of 
$\deff$ in (m19) it follows that
\begin{eqnarray}
\deff\geq 1/p_{\maxi}
\ .
\label{a16}
\end{eqnarray}
 
Upon restricting the integration domain in (\ref{a6})
to $x\geq \tilde a$ and observing (m10) one finds that
\begin{eqnarray}
\int_{\tilde a}^{a_{\maxi}}dx \ w(x)\, 
\frac{1}{1+y(\tilde a)\,(\tilde a-x)}
\leq 1
\ .
\label{a17}
\end{eqnarray}
Exploiting (\ref{a9}) and
performing the integration yields
\begin{eqnarray}
\frac{\tilde w  }{y(\tilde a)}\,
\ln\left(\frac{1}{1-y(\tilde a)\,(a_{\maxi}-\tilde a)}\right)
\leq 1 \ .
\label{a18}
\end{eqnarray}
With (\ref{a15}) it follows that
\begin{eqnarray}
\frac{\tilde w  }{y(\tilde a)}\,\ln(p_{\maxi}N) \leq 1
\label{a19}
\end{eqnarray}
and with (\ref{a11}) that
\begin{eqnarray}
\nu_{\tilde a}\,\frac{\ln(p_{\maxi}N)}{y(\tilde a)\,(a_{\maxi}-\tilde a)} 
\leq 1
\label{a20}
\end{eqnarray}
Taking into account (\ref{a3}) implies
\begin{eqnarray}
\nu_{\tilde a}\ln(p_{\maxi}N)
\leq y(\tilde a)\,(a_{\maxi}-\tilde a)
 \ .
\label{a21}
\end{eqnarray}
Since $p_n\geq 0$ according to (m10),
we can infer from (\ref{a15}) that
$y(\tilde a)\,(a_{\maxi}-\tilde a)\leq 1$ and hence
\begin{eqnarray}
\ln(p_{\maxi}N) \leq 1/\nu_{\tilde a} \ .
\label{a22}
\end{eqnarray}
Utilizing that $\alpha,\,\beta\in\RR$ satisfy 
$\alpha\leq\beta$ if and only if $e^\alpha\leq e^\beta$
it follows that
\begin{eqnarray}
p_{\maxi}\leq \frac{1}{N}\exp\left\{1/\nu_{\tilde a}\right\}
\ .
\label{a23}
\end{eqnarray}
Taking into account (\ref{a5})
one finally recovers the second 
inequality in (\ref{a13}) in the special
case $a=\tilde a$.

Our next observation is that $p_{\maxi}$
in (\ref{a15}), considered as a function 
of $a$, increases monotonically within 
the domain (\ref{a2}).
The derivation of this property from (m10)
and (m11) is straightforward but quite
lengthy, hence the detailed calculations 
will be provided in a separate publication
[1].
A heuristic argument in support of this
property is as follows:
Instead of considering $\tilde a$ as
being fixed via (\ref{a5}), we temporarily
consider $\tilde a$ as variable, but still 
non-negative and so that the right hand side
in (\ref{a5}) is a lower bound for the left hand side.
In other words, $\tilde a$ may now be smaller
than in the case when the identity in (\ref{a5})
applies.
Repeating the same line of reasoning as in
(\ref{a11}) and (\ref{a17})-(\ref{a23}),
one readily finds that the right hand 
side in (\ref{a23}) indeed decreases upon 
decreasing $\tilde a$.
While this argument is strictly speaking restricted
to $\tilde a$ values, to which the approximation 
(\ref{a9}) applies (but (\ref{a10}) is {\em not} required),
the same conclusion can also be derived
without invoking any further assumption 
[1].

Taking for granted that $p_{\maxi}$ is a 
monotonically increasing function of $a$ 
within the domain (\ref{a2}),
it follows with (\ref{a16}) that $\deff$ 
increases upon decreasing $a$.
Given the second inequality in (\ref{a13}) 
has already been verified for $a=\tilde a$,
we can conclude that the same inequality 
must be fulfilled for all $a\in[0,\tilde a]$.

%%%%%%%%%%%%%%%%%%%%%%%%%%%%%%S
\section{Quantitative typicality estimates}
\label{s2}
This section provides the derivation of
the quantitative probabilistic statement 
below Eq. (m20).

Eqs. (m16)-(m19) with
$B=\id$ imply
$[\langle\phi|\phi\rangle]_c = 1$
and 
$[\left(\langle\phi|\phi\rangle- 1 \right)^2]_c \leq\deff^{-1}$.
With the help of Chebyshev's inequality one thus
can infer that
\begin{eqnarray}
& & 
\pr\left(
| \langle\phi|\phi\rangle - 1 |\leq \delta
\right)
\geq 1-\delta \ ,
\label{b1}
\\
& & 
\delta :=  \deff^{-1/3}=[\pu]^{1/3}
\ ,
\label{b2}
\end{eqnarray}
where the left hand side in (\ref{b1})
denotes the probability
that $|\langle\phi|\phi\rangle -1 |\leq \delta$
for a random vector $|\phi\rangle$, 
sampled according to (m13) and (m14).

Similarly, Eqs. (m12), (m15), and (m16) 
with $B=A$ imply that $\mu_{\! A}= a$
and (m8), (m17)-(m19) that
$\sigma_A^2\leq \deff^{-1}$.
Chebyshev's inequality thus yields
\begin{eqnarray}
\pr\left(
| \langle\phi|A|\phi\rangle - a |\leq \delta
\right)
\geq 1-\delta \ .
\label{b3}
\end{eqnarray}
Likewise, by choosing $B=\overline{A_t}$ 
and observing that $\norm{\overline{A_t}}\leq\norm{A}$, 
one obtains
\begin{eqnarray}
\pr\left(
| \langle\phi|
\overline{A_t}
|\phi\rangle - \tr\{\rho \overline{A_t}\} | \leq \delta
\right)
\geq 1- \delta \ .
\label{b4}
\end{eqnarray}

As said in the main paper, 
$S_\phi$ from (m6)
satisfies $[S_\phi]_c\leq 2/\deff$
and $S_\phi\geq 0$.
We thus can invoke Markov's 
inequality to infer
$\pr( S_\phi \leq \delta)\geq 1 - 2 \delta^2$.
Focusing on cases with $\deff \geq 8$,
or equivalently (see (\ref{b2}))
\begin{eqnarray}
\delta \leq 1/2 \ ,
\label{b5}
\end{eqnarray}
it follows that
\begin{eqnarray}
\pr\left( S_\phi \leq \delta
\right)
\geq 1 - \delta \ .
\label{b6}
\end{eqnarray}

Rewriting (m14) and (m20) as
\begin{eqnarray}
|\psi\rangle = \frac{1}{\sqrt{\langle\phi|\phi\rangle}}\, 
|\phi\rangle
\ ,
\label{b9}
\end{eqnarray}
we will tacitly consider $|\psi\rangle$ as a
function of $|\phi\rangle$ from now on.
With the definition
\begin{eqnarray}
q(\phi):=|1-\langle\phi|\phi\rangle|
\label{b7}
\end{eqnarray}
we thus can rewrite
$|\langle\psi|A|\psi\rangle - \langle\phi|A|\phi\rangle|$
as $q(\phi) |\langle\psi|A|\psi\rangle|$.
Exploiting the triangle inequality we can conclude that
\begin{eqnarray}
|\langle\psi|A|\psi\rangle -a| 
\leq 
q(\phi)  |\langle\psi|A|\psi\rangle|
+
|\langle\phi|A|\phi\rangle -a| 
\ .
\label{b10}
\end{eqnarray}
Since $|\langle\psi|A|\psi\rangle|\leq 1$ 
according to (m8),
this yields
\begin{eqnarray}
|\langle\psi|A|\psi\rangle -a| 
\leq 
q(\phi) +
|\langle\phi|A|\phi\rangle -a| 
\ .
\label{b11}
\end{eqnarray}

Due to (\ref{b3}) 
the probability that 
$|\langle\phi|A|\phi\rangle -a|\leq\delta$
is at least $1-\delta$,
and due to (\ref{b1}), (\ref{b7}) 
the probability that $q(\phi) \leq \delta$
is at least $1-\delta$.
Therefore, the probability that both
$|\langle\phi|A|\phi\rangle -a| \leq \delta$
and 
$q(\phi) \leq \delta$
are simultaneously fulfilled
must be at least $1-2\delta$.
Together with (\ref{b11}) we thus can conclude
that
\begin{eqnarray}
\pr\left(
| \langle\psi|A|\psi\rangle - a |\leq 2 \delta
\right)
\geq 1-2\delta \ .
\label{b12}
\end{eqnarray}
Along similar lines, one can deduce from (\ref{b4}) that
\begin{eqnarray}
\pr\left(
| \langle\psi| \overline{A_t} |\psi\rangle - \tr\{\rho \overline{A_t}\} | 
\leq 2 \delta
\right)
\geq 1- 2\delta \ .
\label{b13}
\end{eqnarray}
Furthermore, one can infer from (m6) 
and (\ref{b9}) that 
\begin{eqnarray}
S_\psi=S_\phi/\langle\phi|\phi\rangle^2
\ .
\label{b14}
\end{eqnarray}
The probability that 
$\langle\phi|\phi\rangle\geq 1/2$ can be 
lower bounded by $1-\delta$
by means of (\ref{b1}) and (\ref{b5}).
With (\ref{b6}) it follows that the probability 
that both $S_\phi \leq \delta$
and 
$1/\langle\phi|\phi\rangle^2\leq 4$
are simultaneously fulfilled
must be at least $1-2\delta$.
Due to (\ref{b14}) we thus can infer
\begin{eqnarray}
\pr\left(S_\psi \leq 4 \delta
\right)
\geq 1-2\delta \ .
\label{b15}
\end{eqnarray}

Finally, we can conclude from
(\ref{b12}), (\ref{b13}), and (\ref{b15})
that the three conditions
$| \langle\psi|A|\psi\rangle - a |\leq 2 \delta$,
$| \langle\psi| \overline{A_t} |\psi\rangle - \tr\{\rho \overline{A_t}\} | \leq 2 \delta$,
and
$S_\psi \leq 4 \delta$
will be simultaneously fulfilled 
with probability $P\geq 1-6\delta$.
Note that this represents a non-trivial
result only for $\delta< 1/6$, hence
the additional condition (\ref{b5})
is redundant.
Since $\deff$ is exponentially 
large in the system's degrees of 
freedom (see below (m19))
it follows that $\delta$ in (\ref{b2})
is exponentially small.
Altogether, we thus recover the announced 
statement below Eq. (m20).

%%%%%%%%%%%%%%%%%%%%%%%%%%%%%%S
\section{Equivalence of thermalization and weak ETH}
\label{s3}
In this section, it is shown that weak ETH (wETH) is
necessary and sufficient for thermalization by means of
a more rigorous but less enlightening reasoning than 
below (m27).

%%%%%%%%%%%%%%%%%%%%%%%%%%%%%%S
\subsection{Thermalization implies  weak ETH}
\label{s31}
As in the main paper (see above (m7)),
the eigenvalues and eigenvectors of $A$ are
denoted as $a_n$ and $|\varphi_n\rangle$, respectively,
and hence
\begin{eqnarray}
A=\sum_{n=1}^N a_n\, |\varphi_n\rangle\langle\varphi_n|
\ .
\label{c1}
\end{eqnarray}
With the help of (m10), (m15), 
and (\ref{c1}) one readily verifies that
\begin{eqnarray}
& & [1+y(a)(a-A)]\, \rho = \rhomic
\ ,
\label{c2}
\end{eqnarray}
where, as in the main paper, $\rhomic:=\id/N$ is the
microcanonical density operator and $\id$
the identity on $\hr$, i.e.
$\id=\sum_{n=1}^N |n\rangle\langle n|$
in terms of the energy basis, or equivalently,
$\id=\sum_{n=1}^N |\varphi_n\rangle\langle\varphi_n|$
in terms of the eigenbasis of $A$.

As in the main paper (see (m22) or (m26)), 
the decisive quantity for thermalization is
\begin{eqnarray}
\Delta (a) :=
\tr\{\rho \overline{A_t}\} - 
\tr\{\rhomic \overline{A_t}\}
=
\tr\{\rho \overline{A_t}\}
\ ,
\label{c3}
\end{eqnarray}
where $\overline{A_t}$ is given by (m3),
and where the last identity is a 
consequence of (m7).
In the following, we will need a sufficiently 
precise definition of thermalization 
which is at the same time physically 
reasonable.
Our definition is as follows:
If $|\Delta(a)|$ is smaller (larger) than 
some threshold value $\epsilon\ll 1$
then we say that the system 
does (does not) exhibit thermalization.
For instance $\epsilon$ may represent 
the experimental resolution limit of
the observable $A$.

On the other hand, the decisive quantity 
for wETH (see below Eq. (m27)) is 
\begin{eqnarray}
Q := 
\frac{1}{N}\sum_{n=1}^N (A_{nn})^2
\ ,
\label{c4}
\end{eqnarray}
where $A_{nn}:=\langle n|A|n\rangle$.
Our objective is to show that 
thermalization implies wETH in the 
sense that a small  value of $\Delta (a)$ 
implies a small value of $Q$.

To this end, we temporarily omit
the argument $a$ of $y(a)$
and rewrite (\ref{c2}) as
\begin{eqnarray}
\left[1+a\,y\right] \, \rho = \rhomic +y\, A\,\rho
\ .
\label{c5}
\end{eqnarray}

Multiplying (\ref{c3}) by $[1+ay]$ and exploiting
(\ref{c5}) yields
\begin{eqnarray}
[1+ay]\,\Delta (a) = \tr\{\rhomic \overline{A_t}\}
+ y\,\tr\{A\rho \overline{A_t}\}
\ .
\label{c6}
\end{eqnarray}
Similarly as in (\ref{c3}), the first term on the 
right hand side of (\ref{c6}) is zero.
Multiplying (\ref{c6}) once more by $[1+ay]$ 
and exploiting (\ref{c5}) thus yields
\begin{eqnarray}
[1+ay]^2\,\Delta (a) = y\, \tr\{A \rhomic \overline{A_t}\}
+ y^2\,\tr\{A^2\rho \overline{A_t}\}
\ .
\label{c7}
\end{eqnarray}

Evaluating the first trace on the 
right hand side of (\ref{c7}) in term
of the energy basis $|n\rangle$,
and utilizing (m3) and the definition 
of $\rhomic$ below (\ref{c2}), 
one finds that
\begin{eqnarray}
\tr\{A \rhomic \overline{A_t}\}
=
\frac{1}{N}\sum_{n=1}^N (A_{nn})^2
\ .
\label{c8}
\end{eqnarray}
Combining (\ref{c4}), (\ref{c7}), and (\ref{c8}) 
thus yields
\begin{eqnarray}
y\, Q & = &  [1+ay]^2\,\Delta (a) - y^2\, R(a)
\ ,
\label{c9}
\\
R(a) & := & \tr\{A^2\rho \overline{A_t}\}
\ .
\label{c10}
\end{eqnarray}
Rewriting $A^2\rho \overline{A_t}$ in (\ref{c10})
as $(A^2\sqrt{\rho})\,(\sqrt{\rho} \overline{A_t})$,
where $\sqrt{\rho}$ is defined below (m15),
and exploiting the Cauchy-Schwarz inequality
(see also footnote [41] in the main paper)
implies
\begin{eqnarray}
|R(a)|^2\leq  \tr\{A^4\rho\}\, 
 \tr\{\rho\, (\overline{A_t})^2\}
\ .
\label{c11}
\end{eqnarray}
Evaluating the first trace by means of the eigenbasis
of $A$ yields $\tr\{A^4\rho\}\leq \norm{A}^4\tr \rho$.
In view of (m8) and $\tr\rho=1$ we thus obtain
$\tr\{A^4\rho\}\leq 1$.
Likewise, one finds for the
last factor in (\ref{c11}) that
$\tr\{\rho\, (\overline{A_t})^2\}\leq 1$
and thus
\begin{eqnarray}
|R(a)|\leq  1
\ .
\label{c12}
\end{eqnarray}

From (m8) and (m12) one can infer  
that all admitted $a$ values must satisfy 
$|a|\leq 1$ and hence $|1+ay|\leq 1+|y|$.
Together with (\ref{c9}), (\ref{c12}) 
and reinstalling the argument $a$ of $y(a)$
we thus obtain
\begin{eqnarray}
|y(a)|\, Q & \leq &  (1+|y(a)|)^2\,|\Delta (a)| + |y(a)|^2
\ .
\label{c13}
\end{eqnarray}

Let us now assume that 
the system thermalizes in the sense 
that $|\Delta (a)|\leq\epsilon$ for some
$\epsilon\ll 1$,
see below (\ref{c3}).
More precisely, we only need 
the weaker assumption
that $|\Delta (a)|\leq\epsilon$
is fulfilled at least for 
one $a$ value with the property 
that $|y(a)|\ll 1$ and $|y(a)|\gg\epsilon$
(for instance $y(a)=\sqrt{\epsilon}$;
the existence of such an $a$ 
value is quite plausible
in view of the expansion (m24);
a rigorous justification follows by
observing that $y(a)$ is a continuous
function of $a$, that $y(0)=0$, and 
that $y(a)\to\infty$ for $a\to a_{\maxi}$,
as can be deduced from the
discussion of (m9)-(m12)
in the main paper, 
see also [1].

Upon dividing  (\ref{c13}) by $|y(a)|$ 
and exploiting that $|\Delta (a)| \ll |y(a)| \ll 1$
we finally obtain
\begin{eqnarray}
Q & \ll  &  1
\ .
\label{c14}
\end{eqnarray}
In other words, we have achieved the 
objective stated below (\ref{c4}).
\\[0.5cm]

%%%%%%%%%%%%%%%%%%%%%%%%%%%%%%S
\subsection{Weak ETH implies thermalization}
\label{s32}
Our goal is to show that a small value of $Q$ 
in (\ref{c4}) implies that $\Delta (a)$ in (\ref{c3}) 
is small.

Analogously as in (\ref{c10}), (\ref{c11}) one
can conclude from (\ref{c3}) that
\begin{eqnarray}
|\Delta (a)|^2\leq \tr\{\rho^2\}\, \tr\{(\overline{A_t})^2\}
\label{d1}
\end{eqnarray}
and with (m3), (\ref{c4}) that
\begin{eqnarray}
|\Delta (a)|& \leq & \sqrt{Q \, T(a)}
\ ,
\label{d2}
\\
T(a) & := & N\, \tr\{\rho^2\}
\ .
\label{d3}
\end{eqnarray}
With (\ref{a14}) we obtain
\begin{eqnarray}
T(a) \leq N\, p_{\maxi}
\ .
\label{d4}
\end{eqnarray}

In the remainder of this subsection, 
$Q\ll 1$ is taken for granted, and we
employ the same setup as in 
Sect. \ref{s1}.
However, (\ref{a5}) and (\ref{a10})
are now replaced by
\begin{eqnarray}
\nu_{\tilde a} & =  & 2/\ln(1/Q)\ ,
\label{d5}
\\
\tilde w &  \gg  & 2/\ln(1/Q)
\ .
\label{d6}
\end{eqnarray}
One readily verifies that (\ref{a12}) 
and (\ref{a23}) still remain true.
By inserting (\ref{d5}) into (\ref{a23}) 
it follows that
\begin{eqnarray}
N\, p_{\maxi} \leq 1/\sqrt{Q}
\ .
\label{d7}
\end{eqnarray}
Due to the same arguments as below
(\ref{a23}), the latter bound applies
for all $a$ values with the possible exception
of very small neighborhoods of $a_{\maxi}$
and of $a_{\mini}$.

Introducing (\ref{d7}) into (\ref{d4}) and (\ref{d2})
finally yields
\begin{eqnarray}
|\Delta (a)| \leq Q^{1/4}
\ .
\label{d8}
\end{eqnarray}
Since $Q\ll 1$ we recover the result
announced at the beginning of this subsection.
\\[0.3cm]
\begin{center}
{\bf -----------------------------------------}
\\[1cm]
\end{center}
\begin{description}
\small
\item{ }
[1]
P. Reimann, Dynamical typicality of isolated many-body 
quantum systems, arXiv:1805.07085
\end{description}


\begin{thebibliography}{99}

\bibitem{ale16}
L. D'Alessio, Y. Kafri, A. Polkovnikov, and M. Rigol,
From Quantum Chaos and Eigenstate Thermalization
to Statistical Mechanics and Thermodynamics,
Adv. Phys. {\bf 65}, 239 (2016)

\bibitem{gog16}
C. Gogolin and J. Eisert,
Equilibration, thermalisation, and the emergence
of statistical mechanics in closed quantum systems,
Rep. Prog. Phys. {\bf 79}, 056001 (2016)

\bibitem{deu91}
J. M. Deutsch,
Quantum statistical mechanics in a closed system,
Phys. Rev. A {\bf 43}, 2046 (1991)

\bibitem{sre94}
M. Srednicki, Chaos and quantum thermalization,
Phys. Rev. E {\bf 50}, 888 (1994)

\bibitem{sre96}
M. Srednicki, 
Thermal fluctuations in quantized chaotic systems, 
J. Phys. A {\bf 29}, L75 (1996)

\bibitem{rig08}
M. Rigol, V. Dunjko, and M. Olshanii, 
Thermalization and its mechanism for generic isolated quantum systems, 
Nature (London) {\bf 452}, 854 (2008)

\bibitem{tas96}
H. Tasaki, 
From Quantum Dynamics to the Canonical Distribution: 
General Picture and a Rigorous Example, 
Phys. Rev. Lett. {\bf 80}, 1373 (1998)

\bibitem{rig12}
M. Rigol and M. Srednicki, 
Alternatives to eigenstate thermalization,
Phys. Rev. Lett. {\bf 108}, 110601 (2012)

\bibitem{pal15}
G. De Palma, A. Serafini, V. Giovannetti, and M. Cramer, 
Necessity of Eigenstate Thermalization, 
Phys. Rev. Lett. {\bf 115}, 220401 (2015)

\bibitem{tas16}
H. Tasaki,
Typicality of Thermal Equilibrium and
Thermalization in Isolated Macroscopic 
Quantum Systems,
J. Stat Phys. {\bf 163}, 937 (2016)

\bibitem{bar17}
C. Bartsch and J. Gemmer,
Necessity of eigenstate thermalization for
equilibration towards unique expectation values when starting from generic initial states,
EPL {\bf 118}, 10006 (2017)

\bibitem{shi17}
N. Shiraishi and T. Mori,
Systematic construction of counterexamples to the eigenstate thermalization hypothesis,
Phys. Rev. Lett {\bf 119}, 030601 (2017)

\bibitem{mon17}
R. Mondaini, K. Mallayya, L. F. Santos, and M. Rigol,
Comment on ``Systematic construction of counterexamples to the 
eigenstate thermalization hypothesis'',
arXiv:1711.06279

\bibitem{bar09}
C. Bartsch and J. Gemmer, 
Dynamical Typicality of Quantum Expectation Values, 
Phys. Rev. Lett. {\bf 102}, 110403 (2009)

\bibitem{mul11}
M. P. M\"uller, D. Gross, and J. Eisert,
Concentration of measure for quantum
states with a fixed expectation value,
Commun. Math. Phys. {\bf 303}, 785 (2011)

\bibitem{bri10}
G. Biroli, C. Kollath, and A. M. L\"auchli, 
Effect of Rare Fluctuations on the Thermalization of Isolated Quantum Systems, 
Phys. Rev. Lett. {\bf 105}, 250401 (2010)

\bibitem{ike13}
T. N. Ikeda, Y. Watanabe, and M. Ueda, 
Finite-size scaling analysis of the eigenstate thermalization 
hypothesis in a one-dimensional interacting Bose gas, 
Phys. Rev. E {\bf 87}, 012125 (2013)

\bibitem{beu14}
W. Beugeling, R. Moessner, and M. Haque, 
Finite-size scaling of eigenstate thermalization, 
Phys. Rev. E {\bf 89}, 042112 (2014)

\bibitem{iko17}
E. Iyoda, K. Kaneko, and T. Sagawa,
Fluctuation Theorem for Many-Body Pure Quantum States,
Phys. Rev. Lett. {\bf 119}, 100601 (2017)

\bibitem{yos17}
T. Yoshizawa, E. Iyoda, and T. Sagawa,
Numerical Large Deviation Analysis of Eigenstate 
Thermalization Hypothesis,
Phys. Rev. Lett. {\bf 120}, 200604 (2018)

\bibitem{ber77}
M. V. Berry, 
Regular and irregular semiclassical wave functions,
J. Phys. A {\bf 10}, 2083 (1977)

\bibitem{vor77}
A. Voros,
Asymptotic h-expansions of stationary quantum systems,
Ann. Inst. Henri Poincar\'e {\bf 26}, 343 (1977)

\bibitem{shn74}
A. I. Shnirel'man,
Ergodic properties of eigenfunctions,
Uspekhi Mat. Nauk {\bf 29}, 181 (1974)

\bibitem{col85}
Y. Colin de Verdiere,
Ergodicit\'e et fonctions propres du laplacien,
Commun. Math. Phys., {\bf 102} 497 (1985)

\bibitem{fei85}
M. Feingold, N. Moiseyev, and A. Peres,
Classical limit of quantum chaos,
Chem. Phys. Lett. {\bf 117}, 344 (1985)

\bibitem{hel87}
B. Helffer, A. Martinez, and D. Robert,
Ergodicit\'e et limite semi-classique,
Comm. Math. Phys. {\bf 109}, 313 (1987)

\bibitem{neu29}
J. von Neumann, 
Beweis des Ergodensatzes und des H- Theorems in der neuen Mechanik, 
Z. Phys. 57, {\bf 30} (1929);
English translation: R. Tumulka,
Proof of the Ergodic Theorem and the H-Theorem in Quantum Mechanics,
Eur. Phys. J. H {\bf 35} 201 (2010)

\bibitem{gol10}
S. Goldstein and R. Tumulka,
Long-Time Behavior of Macroscopic Quantum Systems: 
Commentary Accompanying the English Translation of 
John von Neumann's 1929 Article on the Quantum 
Ergodic Theorem,
Eur. Phys. J. H {\bf 35}, 173 (2010)

\bibitem{gol11}
S. Goldstein and R. Tumulka,
On the Approach to Thermal Equilibrium of Macroscopic Quantum Systems,
AIP Conf. Proc. {\bf 1332}, 155 (2011)

\bibitem{rei15}
P. Reimann, 
Generalization of von Neumann's Approach to Thermalization, 
Phys. Rev. Lett. {\bf 115}, 010403 (2015) 

\bibitem{ham18}
R. Hamazaki and M. Ueda,
Atypicality of most few-body observables,
Phys. Rev. Lett. {\bf 120}, 080603 (2018)

\bibitem{rei15b}
P. Reimann,
Eigenstate thermalization: Deutsch's approach and beyond,
New J. Phys. {\bf 17}, 055025 (2015)

\bibitem{rei08}
P. Reimann, 
Foundation of statistical mechanics under 
experimentally realistic conditions,
Phys. Rev. Lett. {\bf 101}, 190403 (2008)

\bibitem{lin09}
N. Linden, S. Popescu, A. J. Short, and A. Winter, 
Quantum mechanical evolution towards equilibrium.
Phys. Rev.  E {\bf 79}, 061103 (2009)

\bibitem{sho11}
A. J. Short, 
Equilibration of quantum systems and subsystems, 
New J. Phys. {\bf 13}, 053009 (2011)

\bibitem{rei12}
P. Reimann and M. Kastner, 
Equilibration of macroscopic quantum systems, 
New J. Phys. {\bf 14}, 043020 (2012)

\bibitem{sho12}
A. J. Short and T. C. Farrelly,	
Quantum equilibration in finite time, 
New J. Phys. {\bf 14}, 013063 (2012)

\bibitem{bal16}
B. N. Balz and P. Reimann, 
Equilibration of isolated many-body quantum systems with 
respect to general distinguishability measures, 
Phys. Rev. E {\bf 93}, 062107 (2016)

\bibitem{f1}
Eq. (\ref{10}) implies
$\sum_{n=1}^N [1+y(a)\,(a-a_n)]p_n=1$.
With (\ref{11}) this yields
$1+y(a)[a-\sum_{n=1}^N a_n\, p_n]=1$
and thus (\ref{12}) if
$y(a)\not=0$.
If $y(a)=0$ we recover (\ref{12})
by observing (\ref{7}), (\ref{10}),
and $a=0$ (see below (\ref{11})).

\bibitem{f2}
Eqs. (\ref{13}), (\ref{14})
imply $ \mu_B=\sum_{m,n=1}^N [c_m^\ast c_n]_c\,
\langle\chi_m|B'|\chi_n\rangle$,
where $\mu_B$ is defined in (\ref{16})
and $B':=\rho^{1/2} B \rho^{1/2}$.
Since $[c_m^\ast c_n]_c=\delta_{mn}/N$ 
(see below (\ref{13})), we obtain
$\mu_B=\tr\{B'\} =\tr\{ \rho^{1/2} B \rho^{1/2}\}=
\tr\{ \rho B\}$, proving (\ref{16}).
Upon verifying and exploiting that
$[c^\ast_jc_kc^\ast_mc_n]_c=
(\delta_{jk}\delta_{mn}+\delta_{jn}\delta_{km})/N^2$,
a similar calculation yields (\ref{17}).

\bibitem{f3}
Viewing $\tr\{C^\dagger D\}$
as a scalar product,
the Cauchy-Schwarz inequality reads
$|\tr\{C^\dagger D\}|^2\leq
\tr\{C^\dagger C\}\tr\{D^\dagger D\}$.
Choosing $C=B\rho $ and $D=\rho B$
yields 
$\tr\{C^\dagger C\}=\tr\{D^\dagger D\}=\tr\{\rho^2B^2\}$
and thus
$\tr\{(\rho B)^2\}=\tr\{C^\dagger D\}\leq \tr\{\rho^2B^2\}$.

\bibitem{sup}
See Supplemental Material below
for mathematical details.

\bibitem{f4}
Due to (\ref{7}), 
$\Amic:=\tr\{\rhomic A\}=\tr\{\rhomic \overline{A_t}\}$ 
happens to vanish,
but for the sake of 
formal clarity, we keep
writing $\Amic$ and
$\tr\{\rhomic \overline{A_t}\}$,
e.g., in (\ref{22}) and (\ref{26}).
For the same reason,
the average of the $A_{nn}$'s 
(over all $n=1,...,N$) happens 
to be zero.

\bibitem{barPC}
C. Bartsch, personal communication.

\bibitem{ste14}
R. Steinigeweg, A. Khodja, H. Niemeyer, 
C. Gogolin, and J. Gemmer,
Pushing the limits of the eigenstate thermalization
hypothesis towards mesoscopic quantum systems,
Phys. Rev. Lett. {\bf 112}, 130403 (2014)

\bibitem{per84}
A. Peres, 
Ergodicity and mixing in quantum theory,
Phys. Rev. A {\bf 30}, 504 (1984)

\end{thebibliography}
\end{document}